%% file: conference_101719.tex
\documentclass[conference]{IEEEtran}
\IEEEoverridecommandlockouts
\usepackage{cite}
\usepackage{amsmath,amssymb,amsfonts}
\usepackage{algorithmic}
\usepackage{graphicx}
\usepackage{textcomp}
\usepackage{xcolor}
\usepackage{ifthen}
\usepackage{enumitem}
\usepackage{hyperref}
\usepackage{svg}
\usepackage{svg-extract}
\usepackage{balance}  
\def\BibTeX{{\rm B\kern-.05em{\sc i\kern-.025em b}\kern-.08em
    T\kern-.1667em\lower.7ex\hbox{E}\kern-.125emX}}
    
\input{memosetup}
\begin{document}

\title{Adaptive Autonomy in Human-on-the-Loop Vision-Based Robotics Systems
}


\author{\IEEEauthorblockN{Sophia Abraham\IEEEauthorrefmark{1}, Zachariah Carmichael\IEEEauthorrefmark{1}, Sreya Banerjee\IEEEauthorrefmark{1}, Rosaura VidalMata\IEEEauthorrefmark{1},\\ Ankit Agrawal\IEEEauthorrefmark{2}, Md Nafee Al Islam\IEEEauthorrefmark{2}, Walter Scheirer\IEEEauthorrefmark{1}, Jane Cleland-Huang\IEEEauthorrefmark{2}}
\IEEEauthorblockA{\IEEEauthorrefmark{1}Computer Vision Research Lab\\\IEEEauthorrefmark{2}DroneResponse Lab\\Department of Computer Science and Engineering \\
\textit{University of Notre Dame}\\
South Bend, USA \\
\{sabraha2, zcarmich, Sreya.Banerjee.9, rvidalma,  aagrawa2, mislam2, wscheire, JaneHuang\}@nd.edu}
}

\maketitle

\begin{abstract}
Computer vision approaches are widely used by autonomous robotic systems to sense the world around them and to guide their decision making as they perform diverse tasks such as collision avoidance, search and rescue, and object manipulation. High accuracy is critical, particularly for Human-on-the-loop (HoTL) systems where decisions are made autonomously by the system, and humans play only a supervisory role.  Failures of the vision model can lead to erroneous decisions with potentially life or death consequences. In this paper, we propose a solution based upon adaptive autonomy levels, whereby the system detects loss of reliability of these models and responds by temporarily lowering its own autonomy levels and increasing engagement of the human in the decision-making process. Our solution is applicable for vision-based tasks in which humans have time to react and provide guidance. When implemented, our approach would estimate the reliability of the vision task by considering uncertainty in its model, and by performing covariate analysis to determine when the current operating environment is ill-matched to the model's training data. We provide examples from DroneResponse, in which small Unmanned Aerial Systems are deployed for Emergency Response missions, and show how the vision model's reliability would be used in addition to confidence scores to drive and specify the behavior and adaptation of the system's autonomy. This workshop paper outlines our proposed approach and describes open challenges at the intersection of Computer Vision and Software Engineering for the safe and reliable deployment of vision models in the decision making of autonomous systems.
\end{abstract} 


\input{1_Introduction}

\input{2_RelatedWork}


\input{4_ProposedFramework}

\input{7_SoftwareEngineering}

\input{6_OpenChallenges}

\input{8_Conclusions}
\section*{Acknowledgment}
This work was partially funded by the US National Science Foundation under grant CNS:1931962. We thank undergraduate students Soumya Abraham and Patrick Soga and graduate student Eric Tsai for helping to develop the weather detection models. We further thank undergraduate students Mike Prieto, Luke Siela, Ben Merrick, Bridget Hart, and Joseph DelleDonne for developing the UI and streaming capabilities to support onboard vision in DroneResponse.

\balance{}
\bibliographystyle{IEEEtran}
\bibliography{IEEEabrv,bibby}



\end{document}

%% file: memosetup.tex
\newboolean{showcomments}
\setboolean{showcomments}{true} 
\ifthenelse{\boolean{showcomments}}
  {
   
  }



%% file: 1_Introduction.tex
\section{Introduction}
Computer Vision (CV) models are broadly utilized within autonomous systems. Examples include driving systems, factory-floor robots, and small Unmanned Aerial Systems (sUAS) deployed for emergency response missions. CV is essential to these applications as it provides critical information about the environment in which the system is operating, and this information is used to support autonomous decision-making. However, CV systems are not entirely reliable and can incorrectly identify, or fail to identify, objects in the real world, leading to incorrect autonomous decisions.
One notable example of CV failure is Uber's Self-Driving car accident on March 19, 2018, which resulted in the fatality of a woman whilst the vehicle was running in autonomous mode with a human driver as a backup. The recorded telemetry showed the vision system had detected and classified the woman six seconds before the crash as an \textit{unknown object}, then as a \textit{vehicle}, and finally as a \textit{bicycle}, resulting in varied predictive responses based on the car's inbuilt autonomy logic. The system finally recognized the need for emergency breaking 1.3 seconds prior to the impact, but that was too late. Uber stated that emergency braking maneuvers were not enabled in these circumstances to reduce ``erratic vehicle behavior'', and furthermore, that the system was not designed to alert the operator. 
A post-mortem analysis identified contributing causes as dark clothing on the pedestrian, lack of side reflectors on the bicycle, front/rear reflectors perpendicular to the path of the vehicle, and no roadway lighting at the location of the incident~\cite{Prelim:2018}.

These types of CV failures have multiple root causes, many of which are introduced whilst training the CV models.  For example, data bias may be introduced by imbalanced data, as in the Uber case, or as human-introduced bias (such as racial biases reflected in recidivism data, or gender biases reflected in census incomes~\cite{Dua:2019}). Bias can cause a model to under-perform in certain circumstances. For example, Wilson et al,~\cite{wilson2019predictive} reported that state-of-the-art object detection systems return poorer performance when detecting pedestrians with darker skin tones regardless of the time of day or whether the person is occluded. When studying the data used to train such models, researchers found that there were about 3.5 times as many samples of people with lighter skin tones than those of people with darker skin. As these kinds of problems are prevalent across almost all current CV models, software intensive systems that leverage CV models must be developed defensively in order to mitigate CV-induced risks.

\begin{figure}[t]
  \centering
  \includegraphics[width=1\columnwidth]{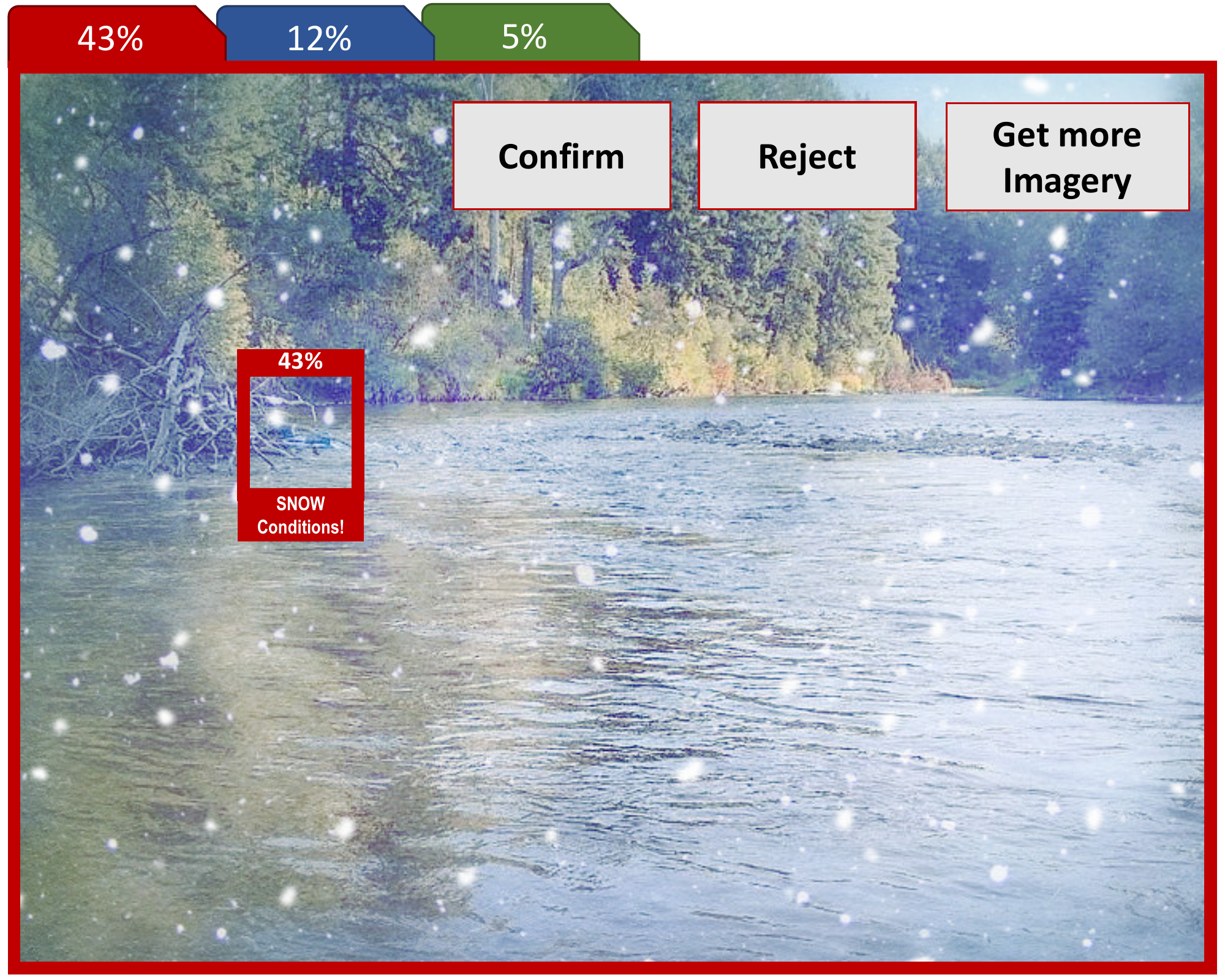} 
  \caption{The sUAS detects a victim with a medium level of confidence and low certainty. It requests confirmation from the human operator before raising a victim-found alert and aborting its search by switching to `tracking' mode. The human can confirm, reject, or request additional imagery. One reason for low confidence is that the training data lacks snowy weather examples. }
   \label{fig:rivervictim}
\end{figure}

\emph{Human-on-the-loop} (HoTL) systems are empowered to make and enact their own decisions~\cite{fischer2017loop} with humans performing only a supervisory role~\cite{nahavandi2017trusted}. Decisions are supported by the system's knowledge of the environment, which is often acquired using CV. In many CV-based scenarios it is essential for the human supervisor to understand whether the vision model can provide reliable results for a specific vision task within the current environment, and whether the system should be trusted to react autonomously, or whether human input is required.  
 Consider the example shown in Figure~\ref{fig:rivervictim} in which an sUAS has detected a potential victim, streams video, and requests input from the human operator. The system can be designed with varying degrees of autonomy with respect to how the sUAS reacts after detecting a candidate victim. For simplicity's sake in this discussion, we assume that only one sUAS has detected the candidate victim -- in this case with a confidence score of 0.43. At this point, the sUAS can (1) ask the operator what to do, (2) automatically track the victim if its confidence  exceeds a predefined fixed threshold, and then notify the operator of its actions (HoTL), or (3) decide whether it is able to start tracking based on the perceived reliability of the CV model within the current environment. The third option represents \emph{adaptive autonomy} in which the sUAS operates independently when it trusts the information provided by the underlying CV models, and engages the operator in critical decisions when the model becomes unreliable.

The aim of this paper is to explore situations in which CV models may suffer from low reliability, propose techniques for detecting failures, and specify requirements for dynamically adjusting autonomy levels and triggering human intervention when the CV model is unable to perform reliably as illustrated in Fig. \ref{fig:context}. We focus upon scenarios in which humans have time to
react and to provide guidance if needed; however, our approach can also be used to raise alerts in situations where the system is making real-time decisions even though CV reliability is low.
We leave the full implementation of the approach to future work. We draw examples from our own HoTL DroneResponse system which deploys multiple sUAS to support time-critical, emergency response missions~\cite{DBLP:conf/chi/AgrawalABCFHHTK20, DBLP:conf/modre-ws/AgrawalCS20}.

The remainder of the paper is structured as follows. Section~\ref{sec:related} presents related work and lays the foundation for our discussion. Section~\ref{sec:cvautonomy} describes our proposed solution for assessing model reliability through evaluating uncertainty and performing covariate analysis with respect to the current environment. Section~\ref{sec:requirements} describes our software engineering solution for CV-driven decision making and autonomy adaptation, while Section~\ref{sec:Challenges} closes with a discussion of open research challenges, and Section~\ref{sec:conclusions} closes with conclusions.



%% file: 2_RelatedWork.tex
\section{Background Information}
\label{sec:related}
A self-adaptive system is capable of reconfiguring at runtime in response to changes in the system and its environment~\cite{DBLP:conf/icse/FredericksC15}. Adaptations include changes in run-time behavior, often realized through switching modes of operation, or by reconfiguring parameters within a mode.  For example, in our DroneResponse system, an sUAS switches between modes (e.g., Takeoff, Search, Track) in response to external events (e.g., destination reached), and can reconfigure its behavior within a mode (e.g., by changing altitude, or increasing monitoring frequency).
However, in this paper, we focus on a special form of adaptation that occurs as a result of uncertainty in the CV model. This form of adaptation represents a temporary switch from HoTL behavior to HiTL (human-in-the-loop) behavior, when the CV model is perceived as unreliable for performing the current task.   Many Software Engineering researchers have explored the notion of uncertainty and its role in self-adaptation~\cite{5328591} including work on identifying gaps in the training data that might make a CV model less reliable in certain environments~\cite{8831191}. AI systems, and CV ones in particular, can cause erratic behavior if they fail to produce accurate results ~\cite{aiid:48,aiid:60,buolamwiniGenderShadesIntersectional2018,geirhosShortcutLearningDeep2020a}. 

\vspace{3pt}\noindent{\it Model Explainability:}
Much related work has focused on explaining predictions made by AI models, especially those leveraging deep neural networks, which tend to provide little or no human-readable rationale for their internal decisions. Their black-box nature can conceal biases, deficiencies, and dubious correlations, which are especially likely when ``dataset shift'' occurs between the training data and the current image stream~\cite{quionero-candelaDatasetShiftMachine2009}. Explainable AI (XAI) techniques add a layer of transparency to this process, whether it is a heat map of pixel importance~\cite{bachPixelWiseExplanationsNonLinear2015,selvarajuGradCAMVisualExplanations2017}, auxiliary model (an explanation model built for each image)~\cite{lundbergUnifiedApproachInterpreting2017,ribeiroWhyShouldTrust2016}, or using attribution based confidence metrics~\cite{jha2019attribution}. Such methods provide explanations in the form of feature attributions (how much they influence the magnitude of a prediction), similar examples, counterfactuals (examples of changes to an image that would have caused a different decision), rules, and visualizations, thus, introducing further clarity during the model development and deployment phases. For instance, feature attribution can indicate that a model relies too heavily on backgrounds of the images rather than the objects of interest in the foreground. Explainable AI can provide useful insights for testing and improving CV models. For example, it might be found that a CV model fails to detect target objects (e.g., people) in rainy conditions due to partial occlusion of the object. This in turn could lead to modified training data including more rainy scenarios. However, explanations of CV decisions deployed in autonomous systems have limited runtime utility, when agents (e.g., sUAS) must make immediate adaptation decisions.  


\vspace{3pt}\noindent{\it Uncertainty Estimation: } Our approach requires the system to evaluate the reliability of the CV model and adapt autonomous behavior accordingly. Uncertainty can be quantified by considering the confidence with respect to the model, the data, and the physical sensor(s). Many approaches have been explored, including augmenting AI systems with auxiliary confidence-estimating modules~\cite{corbiere2019addressing}, directly calibrating the decision probabilities~\cite{DBLP:conf/icml/GuoPSW17}, and taking Bayesian estimates of certainty~\cite{malinin2018predictive}. The first two techniques involve training a separate component of the system to calibrate decisions as probabilities directly interpretable as confidence scores. The latter adds a supplemental model that estimates the conditional distribution of the model decisions given the system and its data. All techniques output the probability that the system will be reliable, regardless of whether the unreliability is caused by unfamiliar data, noise, or other confounding signals.

\vspace{3pt}\noindent{\it Integrating Humans in the Decision-Making Process: } Finally, researchers have investigated when and where humans should be involved in decision-making processes in order to maximize the benefits of autonomy whilst interjecting human feedback when needed for correctness, safety, regulatory compliance, or other purposes. Examples include runtime supervision of autonomous driving systems for safety purposes~\cite{GIL201921}, training machine learning solutions in health informatics to increase their accuracy~\cite{Holzinger2016}, and human-robot partnerships in machine assembly tasks~\cite{raessa2019humanintheloop}. For example, Cai et al.,~\cite{9139404} proposed an approach in which a robot collects multiple images of an object, classifies them using a trained deep convolutional neural network, and when confidence is low for some subset of the images, it determines whether it should autonomously reposition itself to potentially achieve a better viewpoint or whether  it should request help from a remote operator. This approach is similar to our proposed adaptive autonomy solution.


\begin{figure}[t]
  \centering
  \includegraphics[width=1\columnwidth]{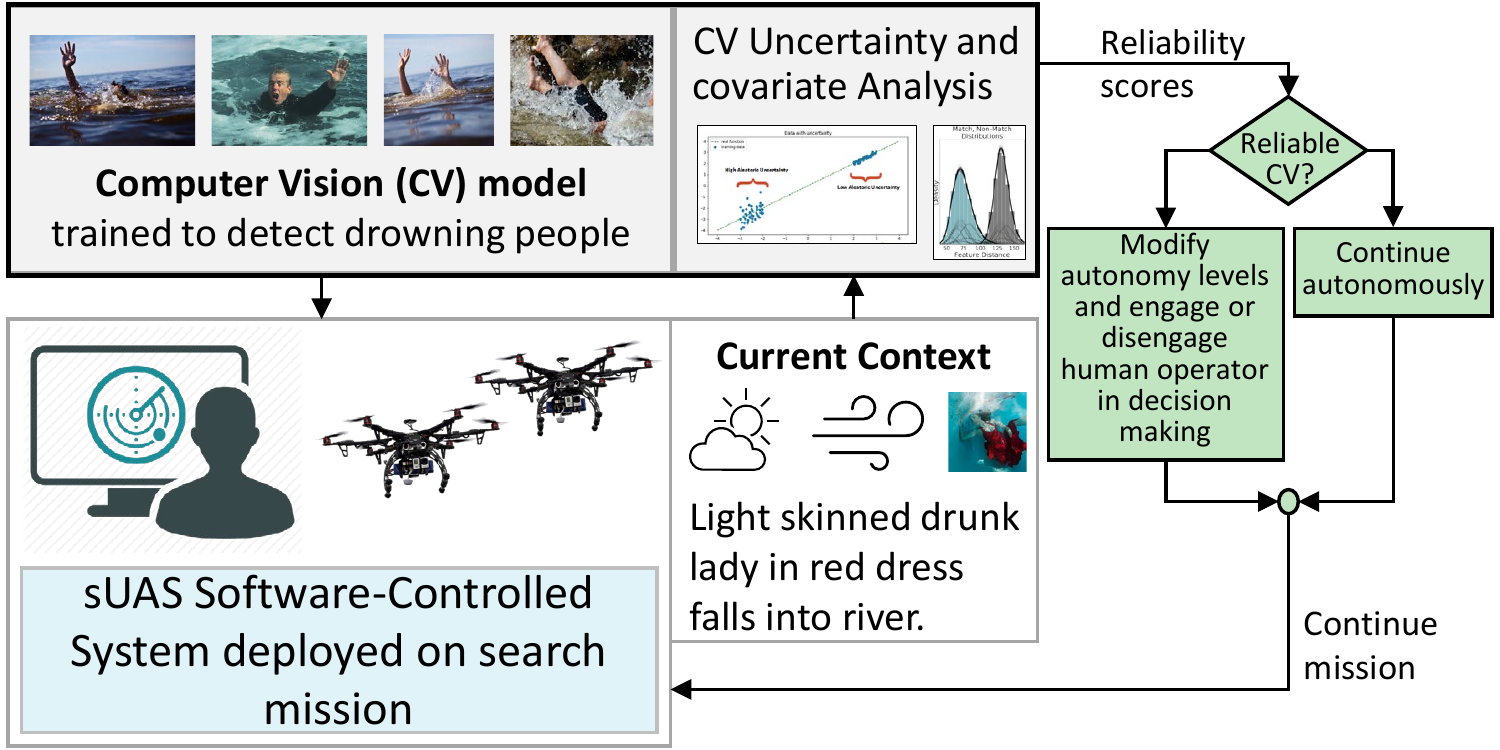} 

  \caption{The vision model is trained on a specific dataset which can introduce bias into the vision-based decision-making. This introduces the critical question of whether the model can be trusted to make a correct decision in the given context.}
   \label{fig:context}
\end{figure}

%% file: 4_ProposedFramework.tex
\section{CV-Supported Autonomy}
\label{sec:cvautonomy}
DroneResponse deploys multiple sUAS to support emergency response missions~\cite{DBLP:conf/chi/AgrawalABCFHHTK20,agrawal2020model,DBLP:conf/modre-ws/AgrawalCS20}. It is designed as a HoTL system empowered to make and enact decisions~\cite{fischer2017loop} supported by its onboard CV. Enabling higher degrees of autonomy is particularly important in DroneResponse missions, where multiple sUAS are deployed simultaneously to perform time-constrained search-and-rescue, surveillance, or delivery tasks.

\subsection{Estimating Loss of Reliability} 
\begin{figure}[t!]
  \centering
    \includegraphics[width=1\columnwidth]{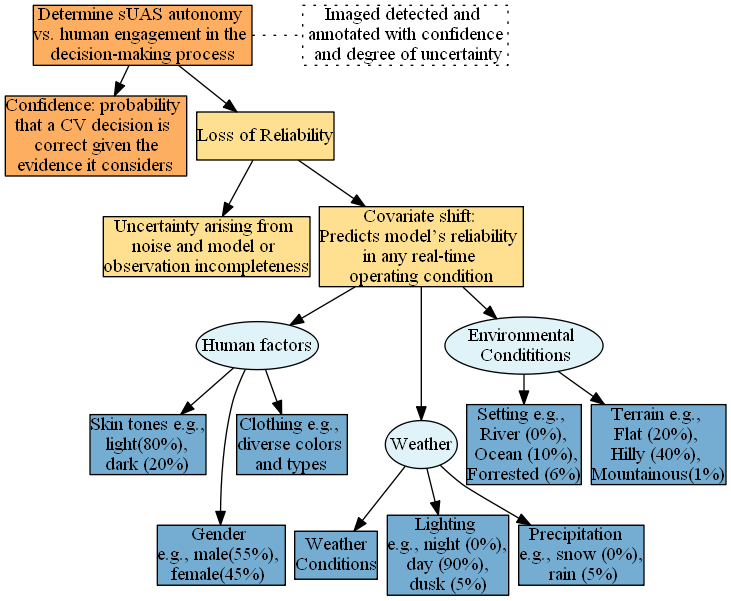}
 \caption{The model's Meta-characteristics are identified for the domain of river search and rescue, that could impact the trustworthiness of the Vision Model. Numbers in the leaf nodes depict the percentage of training samples from the dataset that match each of the meta-tags. Note: These numbers are not derived from actual datasets and are used for illustrative purposes only.}
   \label{fig:matching}
\end{figure}

 Our approach determines the trust that the sUAS system should place in CV predictions based upon confidence and reliability of the underlying model, where confidence is defined as the probability that a CV decision is correct given the evidence it considers, and reliability is estimated using notions of {\it uncertainty} arising from noise and model or observation incompleteness~\cite{pouget2016confidence}. Finally, covariate shift analysis is performed to determine the model's performance under multiple continuous covariates to allow us to estimate the model's reliability in any real-time operating condition. These covariates can be specified as any attributes that may affect the efficacy of the vision model within the given operating domain, such as the attributes depicted in Fig.~\ref{fig:matching} for the DroneResponse river-rescue scenario.

\subsection{Modeling and Detecting Unreliability}
\label{sec:Unreliability}
Our approach utilizes two different techniques for estimating the loss of reliability. The first, is based on the formal notion of uncertainty, derived using a Bayesian Belief Network (BBN), while the second is based on an estimation of the covariate shift between the current scene and the data used to train the model. We discuss each of these approaches.

\vspace{3pt}\noindent{\it Uncertainty in CV Models:}
Uncertainty is typically estimated using Bayesian surrogate estimators, that enable the CV algorithm to infer its own degree of certainty, represented  as one or more probability scores. We adopt state-of-the-art techniques in uncertainty estimation. Loquercio et al., ~\cite{LoquercioUncertainty2020} proposed the use of BBNs and Monte-Carlo sampling to derive uncertainty from both the data and model. Specifically, data noise, arising from the sensor, is assumed to follow a normal distribution based on known noise characteristics. In our sUAS system, these characteristics could include glare, excess vibration caused by the sUAS and/or the camera mount, physical occlusion of objects to be detected, or general image noise.
Uncertainty among the parameters is estimated by Monte Carlo sampling of the parameters with test time dropout -- that is, reducing the population before sampling for the sake of tractability. The model makes several predictions using such subsets, the variance of which is the model certainty. The full system uncertainty is then the total variance of the data uncertainty propagated through the model. In other words, it is the the extent to which the estimation of uncertainty changes between each layer of the CV model, from that induced by sensor noise up to the model parameters of the output layer.

We integrate this technique as a module into our proposed framework, providing a streaming confidence interval in the range $[0,1]$, which comprises uncertainty estimates from both the camera and the vision model, accounting for noise and incompleteness. As a scaffold, we add a hysteresis band that filters transient noise, making the algorithm less sensitive to minor fluctuations which would otherwise produce false positives or negatives. The video stream is transformed into semantic data representing  {\it confident} (no intervention), {\it uncertain} (possible intervention), and {\it no confidence} (intervention required). These parameters can be tuned using field data and adjusted according to intervention budgets. In the case of DroneResponse they would be adjusted based upon a combination of safety factors, event occurrence, and human resources.
For example, if multiple sUAS raised alerts at the same time, and the human operator is unable to process all of them, then the events must be prioritized for intervention, while other sUAS make their best judgments until the operator is able to review, and then confirm or refute their decisions.


\vspace{3pt}\noindent{\it Covariate Shift Analysis:}
In addition to estimating uncertainty generated by the model, it can be helpful to understand the effects of covariates on the model's performance within an operating context. Covariates refer to the known measured attributes within the data. A situation known as \textit{covariate shift} occurs when the training data differs from the data seen at the time a prediction is made. This can result in incorrect predictions with high levels of confidence~\cite{NEURIPS2019_8558cb40}, a problem which is common in real-time detection models. For example, a CV model trained only in good visibility conditions, might underperform in low visibility. 

We propose the formation of a generative model based on the work of McCurrie et al.~\cite{9304938} to capture the distribution between multiple continuous covariates and model performance and to subsequently inform the decision making process. 

\begin{figure}[t]
  \centering
  \includegraphics[width=1\columnwidth]{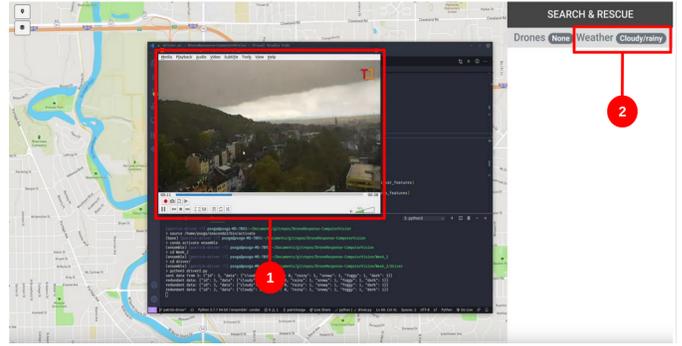} 
  \caption{Our trained weather classifier can take a video stream (1) and tag with weather characteristics (2). Covariate shift can be assessed by comparing the weather distribution for the training data against the current image stream.}
   \label{fig:weatherpoc}
\end{figure}

\begin{figure}[t]
  \centering
  \includegraphics[width=1\columnwidth]{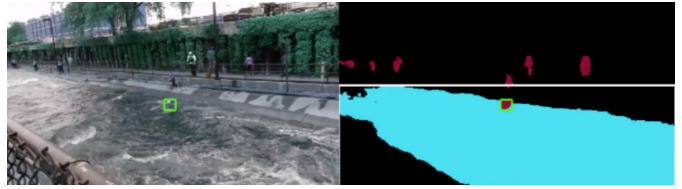} 
  \caption{Our use of the DeepLab~\cite{chen2017deeplab} semantic segmentation model applied to video acquired for a river rescue scenario from the South Bend Fire Department, successfully encoded the pixels corresponding to `river' and `people' from the frame of reference (right panel). Segmentation models could be used to automate the meta-tagging of pertinent covariates.} 
   \label{fig:semanticsegpoc}
\end{figure}

In reference to a person detection model, the generative model could be formulated in three distinct steps:

\begin{enumerate}[leftmargin=*]
     \setlength\itemsep{.4em}
    \item\textit{\it{Data annotation:}} Given a dataset containing $N$ images, we annotate the images with relevant covariates.  While covariates such as age, gender, and race might need to be manually annotated, other covariates such as clothing, angle of view, occlusion, environmental factors, and weather conditions could be extracted with attribute classifiers. Two classifiers that we have developed include a weather classifier and a semantic segmentation model for labeling different parts of the river. Our weather model is built using an ensemble of binary support vector machines trained for pertinent attributes (i.e., light, rain, snow, etc.). It tags each of the $N$ images with corresponding weather and daylight meta-characteristics (Figure \ref{fig:weatherpoc}). Our semantic segmentation model tags data with setting and terrain meta-characteristics (e.g., water, river bank). For this segmentation model, we utilized DeepLab~\cite{chen2017deeplab} for relevant operating covariates (Figure \ref{fig:semanticsegpoc}).

    \item \textit{\it{Estimate pairwise similarities:}}
    The pairwise distance or similarity between the dataset and the image itself is calculated by constructing a matrix in with $N^2$ data points where each row (query) represents the image being sampled, and each column (gallery) represents the collection of images on which the generated model has been formed. Given a data point $(X_k,y_k)$ for $k \in \{1, ..., N^2\}$, $y_k$ represents the similarity between two images and $X_k$ is the vector of all the query and gallery attributes. This also includes a user-defined Boolean attribute which returns `true'  when the query and gallery attributes are matched and `false' when they are not matched.
    \item \textit{\it{Density regression:}} Finally, we estimate the full density of the general distribution of predictive scores over the continuous covariates in order to determine the extent to which the current image from the operating context matches the images in the training set.  We first calculate the conditional true positive rate (TPR): 
     $$TPR(fpr|X)=F_M(F_{\bar{M}}^{-1}(fpr|X)|X)$$
     This provides the TPR at a given false positive rate (fpr) for a precondition $X$, which is the covariate vector mentioned before. Here $F_M$ and $F_{\bar{M}}$ are the cumulative distribution functions (CDF) of the match and non-match distribution. The above relation can be utilized at runtime to estimate the reliability of the model given the covariates of the operating condition.
\end{enumerate}

By extracting and analyzing covariates in this way, and by detecting uncertainty in the CV models with respect to the current CV-task, the system can determine the extent to which it can make autonomous decisions and whether human intervention is required. 

%% file: 7_SoftwareEngineering.tex
\section{Specifying Autonomy Adaptation Requirements}
\label{sec:requirements}
Our framework supports the development of a CV-driven solution with awareness of the model's reliability. Given this self-awareness, we can specify requirements and design a system capable of adapting its own autonomous behavior according to the perceived reliability. Whittle et al.,~\cite{5328591} previously proposed the RELAX language for specifying requirements of self-adaptive systems. RELAX provides the means of describing uncertainty using natural language or Fuzzy branching temporal logic and supports the notion of requirements satisficing in which requirements can be relaxed to address uncertainty. However, in this initial paper, we adopt the simpler EARS (Easy Requirements Specification) notation ~\cite{EARS,DBLP:conf/re/Cleland-HuangV18} which is sufficiently expressive to define adaptive behaviors proposed by our approach. 

We identify human-sUAS interaction points by leveraging our existing meta-model for human-on-the-loop  interactions in multi-agent missions~\cite{DBLP:conf/modre-ws/AgrawalCS20}. In addition to modeling typical human-sUAS interactions, the model includes `probing questions' designed to aid in the discovery and specification of autonomy requirements. The following three questions are particularly pertinent to our discussion of CV-related autonomy and human interventions.
\begin{enumerate}[leftmargin=.9cm]
    \item [{\footnotesize Q1}:] When and where do the agents exhibit {\bf autonomous decision-making behavior?}
    \item [{\footnotesize Q2}:] Under {\bf normal operating conditions}, what decisions should the agent be able to make autonomously? 
    \item [{\footnotesize Q3}:] How is the {\bf autonomy suppressed or increased} at this interaction point? (e.g., modifying the confidence threshold for automatically tracking a potential victim, disabling/enabling the ability to track without permission, ... ) 
\end{enumerate}

We answer these questions in order to specify system level requirements which ultimately must be realized through lower level software requirements and design constraints. Here we focus primarily on design-level requirements which specify rules for switching between autonomous behavior and human-intervention. We err on the side of caution and engage the operator in the decision making process whenever (1) loss of reliability of the CV model exceeds a predefined threshold and/or (2) when the covariate shift indicates that the CV model training data does not provide sufficient coverage for the current environment. We establish internal thresholds for each of these and define \emph{loss\_of\_reliability=TRUE} when either uncertainty exceeds \emph{uncertainty\_threshold} or the covariate analysis returns a score $<$ \emph{covariate\_coverage}.   

We apply the questions (Q1, Q2, Q3) to aid in specifying the requirements for autonomy adaptation of our running example (see Figure~\ref{fig:rivervictim}) in which the sUAS detects a victim and determines what actions should be taken (Q1). To answer Q2 we specify the following autonomy requirements (AR) where $CS$ is the confidence score generated by the CV model.
\begin{enumerate}[leftmargin=.9cm]
    \item [{\footnotesize AR1}:] When the CV model identifies a candidate victim in the river with $CS$ $>=$ \emph{detect\_threshold} and \emph{loss\_of\_reliability=FALSE} then the sUAS autonomously transitions into \emph{tracking} mode and notifies the operator.
    \item [{\footnotesize AR2}:] When the CV model identifies a candidate victim in the river with $CS$ $>=$ \emph{detect\_threshold} and \emph{loss\_of\_reliability=TRUE} then the sUAS temporarily reduces its autonomy level, and raises a high-priority alert requesting permission from the operator to transition into \emph{tracking} mode. 
     \item [{\footnotesize AR3}:]When the CV model identifies a candidate victim in the river with \emph{alert\_threshold} $<$ CS $<$ \emph{detect\_threshold} then the sUAS raises a \emph{low-priority alert} and continues with its current tracking task.
\end{enumerate}
We address Q3 at the design level. Instead of the system limiting the autonomy of the sUAS, we imbue the sUAS with self-awareness so that it detects loss of reliability of the CV model and temporarily overrides its own autonomy to request help in its decision-making ability. 

Returning to our earlier example in which an sUAS detects a candidate victim in the river (cf~Fig. \ref{fig:rivervictim}), if we assume that {\it detect\_threshold} if 0.4, but {\it loss\_of\_reliability}=`true' due to the presence of snow in the operating environment without sufficient representation in the training set, the requirement {\it AR2} is activated and the human is alerted about the potential victim sighting, and engaged in the decision-making process.  In this case, the human might request `get more imagery' triggering the sUAS to reposition itself and stream further imagery, or could reject the sighting and direct the sUAS to continue its search.

This workshop paper focuses on describing how CV confidence, uncertainty, and covariate shift can be used to specify autonomous behavior. Experimental analysis of thresholds, and full integration of our approach within DroneResponse is left for future work.

%% file: 6_OpenChallenges.tex
\section{Open Challenges}
\label{sec:Challenges}
This paper has laid out a practical approach for leveraging CV models within autonomous systems. It describes our proposed framework, which in turn builds upon cutting edge research from both the Computer Vision and Software Engineering communities. To deploy the proposed solution on sUAS operating in potentially unknown environments requires us to address a number of open challenges associated with the CV models themselves, and their safe and reliable adoption within autonomous systems. 

\begin{enumerate}[leftmargin=0cm]
    \setlength\itemsep{.5em}
    \item[] {\bf Challenge \#1, Discovering \& assessing salient covariates:~} As discussed in section~\ref{sec:Unreliability}, identifying covariates is a challenging problem due to the black box nature of neural networks and the vast space of potential covariates. We need the ability to determine which covariates to include in our reliability model, the extent to which they impact performance (individually and as a group) and ways to understand prominent covariates which may not be human-comprehensible. 
    Some related work has been in coercing CV models to discover features in training and visualizing features, e.g.~\cite{Yang_2019_CVPR}. Activation maps can indicate what features invoke larger responses from a model and thus pave the way to covariate discovery.
    The challenge here is to identify the covariates that affect reliability, extract covariates at runtime during an sUAS deployment, and to estimate uncertainty based on these factors.
    \item[] {\bf Challenge \#2, Extracting attributes from image data:~} Most datasets used to train vision models are assembled to support specific tasks.  For example, object detection and recognition datasets such as PASCAL VOC ~\cite{everingham2010pascal}, Imagenet~\cite{russakovsky2015imagenet}, and MS-COCO~\cite{lin2014microsoft} consist of thousands of images with labels for common objects, such as ``person'' with thousands of images representing people. However, none of these particular datasets take into consideration latent characteristics within the image such as light conditions, weather, terrain, or diverse characteristics of the people themselves. As a result, models trained on these datasets can fail due to out-of-distribution inputs~\cite{vidalmata2019bridging,banerjee2019report}. For instance, object detectors such as YOLO~\cite{redmon2018yolov3} trained on the PASCAL dataset, may perform better when localizing people on a bright sunny day than on a rainy or a snowy day since the dataset does not have sufficient representation of images under such weather conditions. When deployed in HoTL environments, their failures can lead to erroneous decisions. However, manually annotating the training set with this information, in order to evaluate covariate shift is time-consuming and difficult.  The challenge is therefore to develop, reliable and automated techniques for meta-tagging covariates in the training set. We provided examples of two techniques for detecting weather conditions and identifying elements of a scene using semantic segmentation.
    \item[] {\bf Challenge \#3,  Capturing real-time context:~} In addition to understanding the distribution of covariates in the training data, we also need to detect relevant covariates within the operating context at runtime. Diverse information sources can be leveraged, such as weather services, onboard sensors, and automated meta-taggers -- potentially using the same classifiers that were applied to the training data. In addition, human input can be elicited to  take advantage of operators' perspectives of the environment.
    \item[]{\bf Challenge \#4: Safety Assurance of CV-Driven systems} Prior research has developed techniques for addressing safety assurance for self-adaptive systems \cite{DBLP:conf/models/ChengCFLM20,DBLP:conf/saso/JahanPGMC19,DBLP:conf/dagstuhl/TrappS11}; however, there is a need for closer integration of the three-way interplay between reliability of state of the art CV models and predictions, the adaptive role of human engagement, and the subsequent creation and generation of safety assurance cases (SAC)~\cite{bloomfield2010safety,holloway2008safety,Hawkins2013,kelly2004goal} which provide evidence for system safety ~\cite{chen-assurancecase,UKMinistryofDefence2017, Graydon2017}. In particular, the safety case must show that thresholds are set at levels that effectively balance agent autonomy with human intervention.
    \item[]{\bf Challenge \#5: Human-Machine Interaction}
    Our proposed approach engages humans-in-the-loop with the aim of increasing accuracy of the CV-driven decision making.  However, prior work has identified different hazards that are introduced when humans place undue trust in the behavior of an autonomous system and as a result, fall out-of-the-loop \cite{endsley2012}, and engage in decision making without sufficient situational awareness. The engagement of humans in supervising and intervening in CV tasks requires careful analysis to explore the tradeoffs of introducing new Human-Computer interactions errors.
\end{enumerate}

%% file: 8_Conclusions.tex
\section{Conclusions}
\label{sec:conclusions}

This paper has presented an adaptive, informal framework for supporting the reliable deployment of CV models in the decision making of autonomous systems and the associated open challenges. It has proposed a framework for determining the reliability of the CV model by estimating the uncertainty of the model and capturing the covariate shift.  The level of autonomy of the system is determined according to the model's reliability. In addition, the paper has described an approach for identifying and specifying uncertainty-driven autonomy requirements driven. We have presented proof-of-concept techniques for parts of our proposed solution, but have  not yet integrated and fully validated all the pieces within our DroneResponse system.